\documentclass[aps,prd,reprint,twocolumn,preprintnumbers,floatfix,nofootinbib]{revtex4-1}
\usepackage[utf8]{inputenc}
\usepackage[dvips]{graphicx}
\usepackage[dvipsnames]{xcolor}
\usepackage{color}
\usepackage{relsize}
\usepackage{graphics}
\usepackage{epstopdf}
\usepackage{hyperref}
\usepackage{mathrsfs}
\usepackage{amssymb}
\usepackage{physics}
\usepackage{booktabs}

\usepackage[normalem]{ ulem }
\usepackage{amsthm}
\usepackage{amsmath}
\usepackage{bm}
\usepackage{cancel}
\usepackage{float}
\usepackage{fontawesome} % for code link icons  
\usepackage[caption=false]{subfig}
\usepackage{tabularx,ragged2e} % new
 \newcolumntype{L}{>{\RaggedRight\arraybackslash}X}

\usepackage{hyperref}

\usepackage{accents}
\newcommand{\der}[2]{\frac{d#1}{d#2}}

\newcommand{\parc}[2]{\frac{\partial#1}{\partial #2}}

\renewcommand{\L}{\mathcal{L}}

\newcommand{\equaref}[1]{Eq.~(\ref{#1})}

\newcommand{\figref}[1]{Fig.~\ref{#1}}

\newcommand{\tabref}[1]{Table~\ref{#1}}

\definecolor{palatd}{RGB}{104, 36, 109}
\definecolor{palatb}{RGB}{0, 56, 168}
\definecolor{palatr}{rgb}{0.745,0.118,0.176}
\newcommand\myshade{80}
\colorlet{mylinkcolor}{palatr}
\colorlet{mycitecolor}{palatb}
\colorlet{myurlcolor}{palatd}

\hypersetup{
  linkcolor  = mylinkcolor!\myshade!black,
  citecolor  = mycitecolor!\myshade!black,
  urlcolor   = myurlcolor!\myshade!black,
  colorlinks = true
}

\graphicspath{{Figures/}}

\begin{document}

\preprint{IPPP/24/12}

\title{Probing the Cosmic Neutrino Background and New Physics with TeV-Scale Astrophysical Neutrinos}
\author{Jack Franklin$^{a}$}
\email{jack.d.franklin@durham.ac.uk}
\author{Ivan Martinez-Soler$^{a}$}
\email{ivan.j.martinez-soler@durham.ac.uk}
\author{Yuber F. Perez-Gonzalez$^{b}$}
\email{yuber.f.perez-gonzalez@durham.ac.uk}
\author{Jessica Turner$^{a}$}
\email{jessica.turner@durham.ac.uk}
\affiliation{$^{a}$ Institute for Particle Physics Phenomenology, Durham University, South Road DH1 3LE, Durham, U.K.}
\affiliation{$^{b}$ Departamento de Física Teórica and Instituto de Física Teórica (IFT) UAM/CSIC, Universidad Autónoma de Madrid, Cantoblanco, 28049 Madrid, Spain}

\begin{abstract}
We use recent evidence of TeV neutrino events from the most significant astrophysical sources detected by IceCube -- NGC 1068, TXS 0506+056, PKS 1424+240-- to constrain the local and global overdensity of relic neutrinos and to explore potential new neutrino self-interactions. Assuming a relic neutrino overdensity, such high-energy neutrinos have travelled considerable distances through a sea of relic neutrinos and could have undergone scattering, altering their observed flux on Earth.  Considering only Standard Model interactions, we constrain the relic overdensity to $\eta \leq 2 \times 10^{14}$ at the 90$\%$ confidence level, assuming the sum of neutrino masses saturates the cosmological bound, $\sum_i m_i = 0.13$ eV. We demonstrate that this limit improves for larger neutrino masses and study how it depends on the scale of the overdensity region. Considering new interactions between TeV-scale neutrinos and relic neutrinos, mediated by a light boson, we probe couplings of approximately $g \sim 10^{-2}$ with current data for a boson mass around the MeV scale. We demonstrate that this limit improves with larger neutrino masses and the scale of the overdensity region. 
\end{abstract}
\maketitle

%%%%%%%%%%%%%%%%%%%%%%%%%%%%%%%%%%%%%%%%%%%%%%%%%%%%%%%%%%%%%%%
\section{Introduction}
%%%%%%%%%%%%%%%%%%%%%%%%%%%%%%%%%%%%%%%%%%%%%%%%%%%%%%%%%%%%%%%
%
The $\Lambda$CDM standard cosmology model predicts a neutrino background akin to the observed cosmic microwave background. 
The discovery of such relic neutrinos would mark a significant breakthrough for cosmology and particle physics, offering a path to unveil fundamental neutrino properties, such as their masses and their nature as Dirac or Majorana particles \cite{Long:2014zva}.
However, detecting these relic neutrinos experimentally has proven to be remarkably challenging. This is primarily because these neutrinos have low momenta, approximately $\sim 0.6~\mu$eV, resulting in only a few possible interaction channels to detect these neutrinos. Nonetheless, there are numerous proposals for relic neutrino detection including capture on radioactive nuclei \cite{Weinberg:1962zza,PTOLEMY:2018jst}, observing the annihilation of ultra-high energy cosmic ray neutrinos with the relic neutrino background using the $Z$-resonance \cite{Eberle:2004ua},  elastic scattering of the relic neutrino wind on a test mass \cite{Stodolsky:1974aq,Opher:1974drq,Duda:2001hd,Domcke:2017aqj,Shvartsman:1982sn,Smith:1983jj,Cabibbo:1982bb}, looking for alterations in atomic deexcitation spectra \cite{Yoshimura:2014hfa},  resonant neutrino capture in accelerator experiments \cite{Bauer:2021uyj} and indirect constraints \cite{McKeen:2018xyz,Arvanitaki:2022oby}. 

These ideas, which may become experimentally feasible in the future, are hindered by anticipated minuscule rates of detected neutrinos; see Ref.~\cite{Bauer:2022lri} for a comprehensive overview of experimental sensitivities. This necessitates either large quantities of detector material or the experimental detection of extremely subtle effects, rendering detection exceptionally challenging.
However, there is a caveat that could potentially alter the anticipated rates in a future facility: while the $\Lambda$CDM model predicts a neutrino density of approximately $\sim 56~\nu/{\rm cm^3}$ for each mass state (and similarly for antineutrinos) \cite{Giunti:2007ry}, there could be an enhancement of the neutrino number density due to gravitational clustering.
Such an overdensity is $ \sim 20\% $ above the average predicted by the $ \Lambda $CDM model. However, it could be much larger depending on whether additional BSM interactions impact neutrinos.
Thus, it may be prudent to first experimentally constrain potential overdensities before attempting a direct detection of the Cosmic Neutrino Background (C$\nu$B).

Constraints on overdensities stem from both theoretical and experimental sides.
On the one hand, clustering large amounts of neutrinos would be prohibited by the Pauli exclusion principle~\cite{Tremaine:1979we,Bauer:2021uyj}. 
Since the standard C$\nu$B is expected to be close to the Pauli limit, any large overdensity would indicate not only the presence of additional interactions, but also that the relic neutrinos would possess a larger temperature to the one expected in standard cosmology.
On the other hand, the KATRIN experiment recently has managed to constrain the local overdensity through a direct search for events consistent with the detection of the C$\nu$B, resulting in a limit of $\sim 10^{11}$ times the average density predicted by cosmology~\cite{KATRIN:2022kkv}.
Another possible way to constrain relic neutrino overdensities consists in determining the effects from scattering of high energy particles with the C$\nu$B, such as cosmic rays~\cite{Ciscar-Monsalvatje:2024tvm}, detecting absorption features in the cosmogenic neutrinos spectrum~\cite{Brdar:2022kpu}, or considering the gravitational effect on Solar System objects~\cite{Tsai:2022jnv}.

A recent analysis of the high-energy events detected by IceCube has unveiled the origins of some of these astrophysical neutrinos. The most significant among the candidate sources is the active galaxy NGC~1068~\cite{IC:NGC1068, PhysRevLett.124.051103}, with a global significance of $4.2\sigma$. Other notable candidate sources with high significance include PKS 1424+240 ($3.7\sigma$)~\cite{Paiano:2017pol} or TXS 0506+056 ($3.5\sigma$)~\cite{IceCube:2018cha}. The discovery of these neutrino point sources presents an opportunity for studying scenarios Beyond the Standard Model (BSM) by exploring deviations in the expected neutrino flux from the source, see e.g.~\cite{Ng:2014, Esteban:2021, Carloni:2022cqz, Rink:2022nvw, Cline:2023tkp, Doring:2023}. This work focuses on a scenario where neutrino self-interacts via Standard Model processes or where a light boson mediates the dominant self-interactions. In the latter case, the strongest constraints on this type of interaction come from Big Bang Nucleosynthesis (BBN)~\cite{Blinov:2019gcj}, which imposes lower bounds on the scalar boson mass of $m_{\phi} > 1.3$~MeV and a coupling $\text{g} \geq 10^{-4}$. For higher masses, additional constraints arise from laboratory measurements~\cite{Berryman:2022hds,Lessa:2007up}. However this case is motivated by the potential alleviation of the $H_0$ and $\sigma_8$ tensions in cosmology by adding BSM neutrino self-interactions \cite{Esteban:2021,Kreisch:2019yzn}. We investigate this scenario by considering self-coupling interactions between relic neutrino and the high-energy neutrino emitted by Active Galactic Nuclei (AGN) observed by IceCube. Our analysis uses the data from the three most significant candidate sources mentioned before. 
Beyond BSM physics, these high-energy neutrinos traverse extensive distances through the C$\nu$B on their way to Earth. In the Standard Model limit, overdensities of these relic neutrinos, which lie along the path traversed by the astrophysical neutrinos, can potentially modify the astrophysical neutrino fluxes through the self-interactions of neutrinos. In this work, we aim to constrain the overdensity of relic neutrinos by analysing neutrino events from the three most significant candidate sources using publicly available data from IceCube.

This work is organized as follows.  
In Sec.~\ref{sec:nu_flux}, we introduce the transport equation that forms the basis of our study.  
Our theoretical framework is presented in Secs.~\ref{sec:overdens} and \ref{sec:self_int}, where we examine two scenarios: the existence of a relic overdensity and the presence of neutrino self-interactions.  
In Sec.~\ref{sec:analys}, we describe the analysis performed for the aforementioned neutrino point sources discovered by IceCube, followed by Sec.~\ref{sec:results}, where we present our findings.  
Finally, in Sec.~\ref{sec:conc}, we summarize our conclusions.  
Additionally, Appendices~\ref{sec:SolvingTransportEq} and \ref{sec:CrossSections} provide further details on the numerical methods used in our study and the derivation of total and differential cross-sections for neutrino-neutrino elastic scattering within the Standard Model.  
We adopt natural units where $\hbar = c = k_{\rm B} = 1$ throughout this work.

%%%%%%%%%%%%%%%%%%%%%%%%%%%%%%%%%%%%%%%%%%%%%%%%%%%%%%%%%%%%%%%
\section{Propagation of Neutrino Fluxes}\label{sec:nu_flux}
%%%%%%%%%%%%%%%%%%%%%%%%%%%%%%%%%%%%%%%%%%%%%%%%%%%%%%%%%%%%%%%

To take into account the effect of scattering with relic neutrinos on the neutrino flux propagating to the Earth, it is necessary to solve a transport equation for the flux of neutrinos with mass state $i$ \cite{Esteban:2021,Bhattacharjee:1999mup},
\begin{widetext}
    \begin{equation}\label{eq:CnuB_transport}
            \frac{\partial \Phi_i (t, E)}{\partial t} = \frac{d}{d E_\nu}\left[ H(t) E_\nu \Phi(t, E)\right]
            -\Phi_i (t,E) \sum_j n_j \sigma_{ij}(E)
            +\sum_{j,k,l} n_k \int_E^\infty dE' \Phi_j (t,E') \frac{d \sigma_{jk\rightarrow il}(E',E)}{dE}\,,
    \end{equation}
\end{widetext}
where $\Phi_i$ denotes the combined flux of neutrinos and anti-neutrinos with mass state $i$, $t$ is the time since the neutrinos were emitted. The first term, containing the Hubble expansion rate $H(t)$, accounts for the energy loss of neutrinos due to the redshift from the expansion of the Universe. The second term is the loss term, which reduces the flux at a particular value of energy according to the interaction rate with the C$\nu$B, which is the product of $n_j$, the number density of mass states $j$, and $\sigma$, the neutrino-neutrino cross-section. In the third term of \equaref{eq:CnuB_transport}, the $j$ state is the incoming neutrino with energy $E'$, the $k$ state is the relic neutrino, $i$ is the outgoing neutrino with the mass state of interest, and $l$ is the other neutrino state produced in the interaction. This term distributes the flux from down-scattering (energy loss of the neutrinos) and up-scattering of the relic neutrino. We solve this differential equation numerically, using a similar method to that in \cite{Esteban:2021}, the details of which can be found in Appendix \ref{sec:SolvingTransportEq}.\newline
The centre-of-mass energy of the interaction of the astrophysical neutrino, with energy $E$, with the relic neutrino of mass $m_j$, assumed to be at rest\footnote{Note that large overdensities due to some BSM interactions might also imply that the relic neutrinos could be relativistic today, enhancing the centre-of-mass energy. However, we refrain from considering such a scenario to keep our discussion as model-independent as possible.}, is
\begin{align}\label{eq:CoM}
    \sqrt{s_j} &= \sqrt{2 m_j E} \notag\\
    &\sim 1.41~{\rm MeV} \left(\frac{m_j}{0.1~{\rm eV}}\right)^{1/2} \left(\frac{E}{10~{\rm TeV}}\right)^{1/2}\,.
\end{align}
For the neutrinos originating from astrophysical sources, we assume an initial power-law (PL) flux, where the flux is parameterised in terms of a normalisation $\Phi_{0}$, at reference energy $E_0$, and a spectral index $\gamma$ such that 
\begin{equation}
\Phi(t=0,E_\nu) = \Phi_0 \left(\frac{E_\nu}{E_0}\right)^{-\gamma}\,.
\end{equation}
We take $E_{0} = 1$~TeV throughout this work. For each source, we assume the fluxes are independent and uncorrelated. We also assume that the neutrino flavours are produced with the initial ratio of 1:2:0 for $\nu_e$:$\nu_\mu$:$\nu_\tau$, which corresponds to pion decays~\cite{Abdullahi:2020}. 
%

%%%%%%%%%%%%%%%%%%%%%%%%%%%%%%%%%%%%%%%%%%%%%%%%%%%%%%%%%%%%%%%
\section{Relic Overdensity}\label{sec:overdens}
%%%%%%%%%%%%%%%%%%%%%%%%%%%%%%%%%%%%%%%%%%%%%%%%%%%%%%%%%%%%%%%

Any deviation from the expected average number density of relic neutrinos, $n_0 \approx 56\, \mathrm{cm}^{-3}$, can be parameterised by $\eta = n/n_0$, where $n$ is the actual number density. 
Simulations of relic neutrinos within the galactic gravitational field generally suggest an overdensity in the range of $\eta_\nu \simeq 1-10$ under conservative scenarios \cite{deSalas:2017wtt,Mertsch:2019qjv,Yoshikawa:2020ehd,Zimmer:2023jbb, Elbers:2023mdr, Elbers:2023wbf}, based on the standard cosmological evolution. Analytic techniques based on kinetic field theory yield similar results \cite{Holm:2023rml}. In Ref.~\cite{Ringwald:2004np},  the relic neutrino gravitational clustering as a byproduct of the clustering of dark matter was investigated. They found relic neutrino number overdensities of the order $1-10^4$. Simulations which assume that the local overdensity is proportional to baryon overdensities found that the neutrino overdensities be estimated, perhaps optimistically, to as large as $\sim 10^{4}-10^{6}$ \cite{Lazauskas:2007da,Faessler:2014bqa}. However, models which invoke new physics beyond the Standard Model can predict large overdensities. For example, Yukawa interactions of ultralight scalars with neutrinos or neutrinos condensates can provide overdensities of $\sim \mathcal{O}(10^{7})-\mathcal{O}(10^{9})$ \cite{Smirnov:2022sfo,Dvali:2016uhn}. 

\begin{figure}[t!]
\centering
    \includegraphics[width=\linewidth]{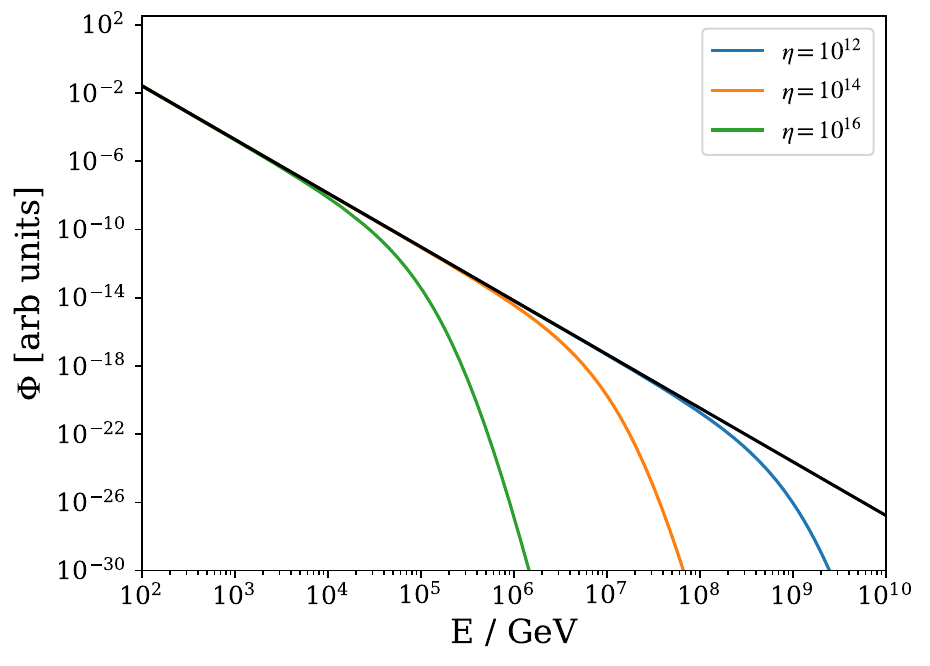}
    \caption{Normalised muon neutrino fluxes from NGC 1068 at IceCube under different scenarios. The black line is the initial flux, taken to be a power law. The spectral index of the initial flux is the best-fit values from our TS analysis, $\gamma=3.15$. Given this initial flux, the other lines are the solutions to the transport equation, with the relevant neutrino number density ratios indicated. The number density is assumed to be constant over the distance between NGC 1068 and Earth, and $m_1$ (the lightest neutrino mass) is $0.034$\,eV. }
    \label{fig:fluxes}
\end{figure}

The final flux at Earth can be found by solving the transport equation of \equaref{eq:CnuB_transport} numerically over the distance that the neutrinos travel through the overdense region, $d_\mathrm{eff}$, assuming a given overdensity, $\eta$.
 For a density that varies in value over spatial coordinates, for example, with gravitational clustering \cite{deSalas:2017wtt}, $d_\mathrm{eff}$ is the radius of a constant density profile that would produce the same number of interactions.
The effect of neutrino self-interactions over the astrophysical neutrino flux can be seen in \figref{fig:fluxes}. Taking NGC 1068 as an example, the initial flux is plotted alongside the final flux for several different relic neutrino densities. The main effect is a suppression of the flux at high energies. Regeneration processes lead to an increased flux at lower energies; however, the influence at lower energies remains minimal due to the rapid decrease of the flux with energy. 
%
%%%%%%%%%%%%%%%%%%%%%%%%%%%%%%%%%%%%%%%%%%%%%%%%

We begin our discussion by considering only SM neutrino interactions, and assume that the number density is the same for all mass states and neutrinos/antineutrinos. We require both total and differential SM cross-sections to solve the transport equations in Eq.~\eqref{eq:CnuB_transport}, which we will briefly outline next. For the observable neutrinos under consideration, which have energy $\lesssim 100\, \mathrm{TeV}$, the centre of mass energy of the interactions (see \equaref{eq:CoM}) are always sufficiently small that the only important processes are $\nu \nu \rightarrow \nu \nu$, $\nu \bar{\nu} \rightarrow \nu \bar{\nu}$, and $\nu \bar{\nu} \rightarrow e^+ e^-$. 

The total cross-section for the SM interactions between the high-energy neutrinos flux from astrophysical sources and relic neutrinos can be split into two contributions: one from the production of neutrino final states only and one from the production of electron-positron pairs:
\begin{equation}
    \sigma_{ij} = \sigma^\nu_{ij} + \sigma^e_{ij}\,,
\end{equation}
where $i$ is the incoming neutrino mass state, and $j$ is the relic neutrino mass state, as before. From this, we find that
\begin{equation}
    \sigma^\nu_{ij} = \frac{G_F^2 (7\delta_{ij}+2)s_j}{3\pi}\,, 
\end{equation}
where $G_F$ is the Fermi constant. This cross-section has been summed over the contributions from relic neutrinos and antineutrinos. For the cross-section with $e^+e^-$ final states, the interaction must be calculated in the mass basis, such that
\begin{equation}
    \sigma^e_{ij} = \sum_\alpha \abs{U_{ei}}^2\abs{U_{ej}}^2 \sigma^e_{\alpha}\,,
\end{equation}
where $\alpha$ is the flavour of the initial state neutrinos in the interaction. 
The production of $e^+e^-$ only occurs with neutrino-antineutrino annihilation, so we can consider the initial flavours to be identical. The total cross-sections for this process are
\begin{equation}
  \sigma_{\alpha}^e = \frac{G^2_F}{3 \pi}
  [s(g_{A}^2+g_{V}^2) + 2 m^2_{e}(g^2_{V} -2g^2_{A})]
  \sqrt{1-\frac{4 m^2_{e}}{s}}\,,
\end{equation}
where $g_{A} = \delta_{\alpha e} + g^{\alpha}_{A}$ and $g_{V} = \delta_{\alpha e} + g^{\alpha}_{V}$ where $g^{\alpha}_{A} = -\frac{1}{2}$ and $g^{\alpha}_{V} = -\frac{1}{2} +2\sin^2\theta_{W}$,
with $\theta_W$ being the weak mixing angle and $m_e$ the electron mass.
Unlike in the case of the total cross-sections, when calculating the differential cross-section in the second term on the RHS of \equaref{eq:CnuB_transport}, we are interested in the kinematics of the final state of the interaction.
In particular, we wish to obtain the differential cross-section in terms of the energy of the outgoing $i$ mass state (anti)neutrino. We find that the differential cross-section for neutrino pair production, taking all processes into account, is
\begin{equation}
    \frac{d\sigma_{jk\rightarrow il}(E^\prime,E)}{dE} = \frac{G_F^2 m_k}{2\pi}\left(A_{ijkl} + B_{ijkl}\left(\frac{E}{E^\prime}\right)^2 \right)\,,
\end{equation}
where we have defined 
\begin{subequations}
    \begin{align}
        A_{ijkl} &= (\delta_{ij}\delta_{kl}+\delta_{ik}\delta_{jl})^2\,,\\
        B_{ijkl} &= (\delta_{jk}\delta_{il}+\delta_{ij}\delta_{kl})^2\,.
    \end{align}
\end{subequations}
We provide details of this calculation in the Appendix \ref{sec:CrossSections} for the interested reader.

From the kinematics of the interactions, we find that the opening angle of the neutrinos produced in these interactions is $O(10^{-6})$ radians over the energy range of interest at IceCube. Whilst this opening angle may produce substantial effects over the large distance between the point sources and Earth, the average opening angle from the production of neutrinos from pions is larger by at least three orders of magnitude. Thus, we can assume that the decrease in flux due to the angular diffusion is counteracted by the same effect occurring in adjacent patches of space.

%%%%%%%%%%%%%%%%%%%%%%%%%%%%%%%%%%%%%%%%%%%%%%%%%%%%%%%%%%%%%%%
\section{Scalar mediated neutrino self-interactions}\label{sec:self_int}
%%%%%%%%%%%%%%%%%%%%%%%%%%%%%%%%%%%%%%%%%%%%%%%%%%%%%%%%%%%%%%%

Treating the SM as a low-energy effective field theory, we can explore the inclusion of a higher-dimensional operator. At dimension five, the Weinberg operator~\cite{Weinberg:1979sa} can explain the smallness of the neutrino masses. At higher dimensions, additional BSM processes can emerge. For example, at dimension six, a coupling between neutrinos and a leptonic scalar with a lepton number charge of -2 can arise through the operator $(LH)(LH)\phi/\Lambda^2$ \cite{deGouvea:2019qaz}. After the electroweak (EW) symmetry breaking, this operator contributes to the Lagrangian as 
\begin{equation}
\mathcal{L} \supset \frac{1}{2} g_{\alpha\beta} \phi \nu_{\alpha}\nu_{\beta}\,,
\end{equation}
where Greek letters denotes flavor indices, $\nu_{\alpha}$ are the left-handed neutrinos, and $\phi$ is a real scalar. While this interaction could induce flavour-changing processes, we focus exclusively on flavour-preserving interactions in this work. We follow the approach outlined in~\cite{Esteban:2021} to incorporate this new interaction into the cross-section calculations.

There are several scenarios in which such interactions have been investigated. For example, in cosmology, these interactions can alter the number of relativistic degrees of freedom ($N_{\text{eff}}$) by heating the neutrino population. Constraints on $N_{\text{eff}}$ from BBN~ \cite{Blinov:2019gcj} impose an upper limit on the mediator mass $m_{\phi} > 1.3$~MeV for couplings $g> 10^{-4}$.  Experiments involving meson and lepton decays also probe these interactions by searching for decay modes involving neutrinos. In this work, we focus specifically on couplings involving tau neutrinos, as their detection is more challenging, leading to weaker coupling strength bounds. Current constraints from measurements of $\tau$ decay rates limit the coupling to $g^2_{\tau\tau} < 0.1$~\cite{Lessa:2007up}.

The coupling between $\tau$ neutrinos is of particular interest for alleviating the $H_0$ and $\sigma_8$ tensions in cosmology, as the preferred regions of parameter space have been ruled out for all other lepton flavours \cite{Esteban:2021}. For $g_{\tau\tau}$, the region of parameter space which results in "moderately interacting neutrinos" (MI$\nu$) \cite{Kreisch:2019yzn}, is still viable. Hence, this coupling will be the focus of our analysis.

%%%%%%%%%%%%%%%%%%%%%%%%%%%%%%%%%%%%%%%%%%%%%%%%%%%%%%%%%%%%%%%
\section{Analysis}\label{sec:analys}
%%%%%%%%%%%%%%%%%%%%%%%%%%%%%%%%%%%%%%%%%%%%%%%%%%%%%%%%%%%%%%%
%
To search for signals of neutrino self-interactions - from both overdensities and non-standard interactions (NSI) - within the IceCube data, we perform an unbinned maximum likelihood test using the SkyLLH python package~\cite{Wolf:2019cfm,IceCube:2023ihk,IceCube:2021mzg}. The likelihood function for $N$ events, with a source flux determined by a set of model parameters $\bm{\theta}$, is
\begin{equation}
    \mathcal{L}(n_s, \bm{\theta}| \mathbf{x},N)  = \prod_{i=1}^N \left(\frac{n_s}{N} f_S(\mathbf{x}_i | \bm{\theta}) + \left(1-\frac{n_s}{N} \right) f_B(\mathbf{x}_i) \right)\,,
\end{equation}
where $\mathbf{x}_i$ are the observables of the event $i$, $n_s$ is the number of events associated with the signal, and $f_S$ and $f_B$ are the signal and background probability distribution functions (pdf), respectively. When performing the analysis, we consider the events from 2012-2018 taken from the public release~\cite{IC:public_release}, and select those within a 15$^\circ$ radius from each source.

To perform a statistical test for a choice of parameters in a given model, we compare the likelihood to that of the null hypothesis, which we take to be the best fit power law flux for the chosen source. We use a log-likelihood ratio as our test-statistic:
\begin{equation}
    \label{eq:TS}
    TS = -2\Delta\log\mathcal{L} = -2 \log\left(\frac{\mathcal{L}(ns, \bm{\theta} | \mathbf{x}_i, N)}{\mathcal{L}_0}\right)\,,
\end{equation}
where we denote the likelihood of the best-fit power law flux hypothesis as $\mathcal{L}_0$. 
The relevant model parameters for the power-law hypothesis are the spectral index $\gamma$, and the location of the source in the sky $\theta_s$. We take fixed values of $\theta_s$ according to the source under consideration. For the overdensity hypothesis we also have the parameter $\eta$, which is the ratio compared to the SM relic neutrino density. When modeling the SI flux, we fix $\eta=1$ and instead have the mass of the scalar mediator $m_\phi$ and the coupling between tau neutrinos $g_{\tau\tau}$ as the free model parameters. For the two latter cases, we may also allow the mass of the lightest neutrino to vary - the other two masses are derived from this value using measured mass splitting values from NuFit 2022 \cite{NuFIT:2022}.

For each realisation of the BSM model parameters ($\eta$ or $(M_\phi, g_{\tau\tau})$), we minimise the test statistic ($TS$) over the spectral index of the initial flux, $\gamma$, and the number of signal events $n_s$. The likelihood of the null hypothesis ($\L_0$) corresponds to the absence of observable signals of neutrino interactions - the scenario where the C$\nu$B density follows the prediction of the $\Lambda$CDM model or where there are no BSM self-interactions - and the neutrino flux is described by the best-fit parameters given in \tabref{Tab:table}.
These values are obtained through a $TS$ analysis comparing the likelihoods between the power-law model and the scenario where all the data corresponds to the background.
\begin{table}[!h]
\begin{center}
\begin{tabular}{||c | c | c||} 
 \hline
 Source & Number of events ($n_{s}$) & Spectral index ($\gamma$)\\ [0.5ex] 
 \hline\hline
 NGC 1068 & 56.5 & 3.15\\ 
 \hline
 PKS 1424+24 & 48.7 & 3.86 \\
 \hline
 TXS 0506+056 & 14.5 & 2.17 \\[1ex] 
 \hline
\end{tabular}
\end{center}
\caption{The best-fit values of the signal event normalisation ($n_{s}$) and the spectral index ($\gamma$) for each of the three sources included in the analysis, assuming SM interactions and no overdensity in the relic neutrino flux.}
\label{Tab:table}
\end{table}

To perform this analysis, it is necessary to solve \eqref{eq:CnuB_transport} for the given model parameters. This results in the flux of neutrinos at IceCube, which can then be used to calculate the signal pdf. 

\subsection{Future Limits}

In order to estimate the sensitivity of the analysis technique in the light of future data, it is necessary to produce mock data on which the analysis can be performed. We do this on a per source basis, on account of the sources being sufficiently separated in the sky that there is minimal probability that an event originating from one source will be within the angular cut taken for the analysis of another source. We also assume that all of the systematics of the IceCube detector will remain the same, which is a decision driven by necessity rather than realism. The creation of mock data can be performed by SkyLLH, using the constructed pdf of the background and a given source pdf. For the latter, we assume the source follows the best-fit power law flux. Once a set of mock data is generated, it is combined with the current data set and the analysis is performed. This process is repeated over a number of different instances, and the expectation value of the test-statistic is approximated as the mean value of $TS$ over all instances of mock data. We found, through a process of trial-and-error, that a set of 500 instances produced reliable values.

%%%%%%%%%%%%%%%%%%%%%%%%%%%%%%%%%%%%%%%%%%%%%%%%
\begin{figure}[t!]
    \centering
        \includegraphics[width=\linewidth]{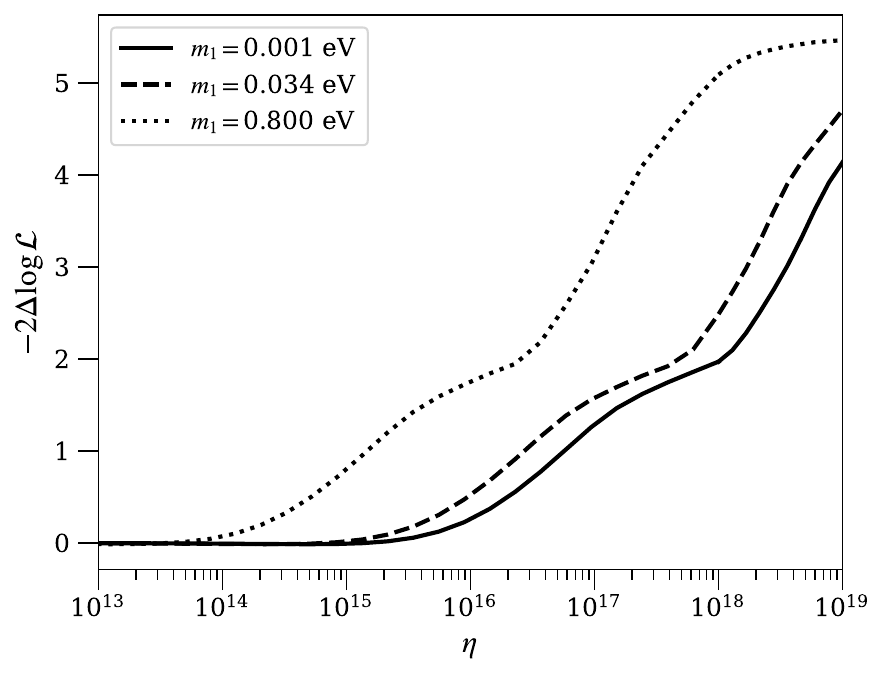}
    \caption{The t-statistic $-2\Delta\log{\L}$ for NGC 1068 as a function of the relic neutrino overabundance. The effect of the value of $m_1$ is also demonstrated by taking different limits for the mass scale, as explained in the text. Here we take $d_{\rm eff} = 14$ Mpc for the radius of the overdense region.}
    \label{fig:line_plot}
\end{figure}
%%%%%%%%%%%%%%%%%%%%%%%%%%%%%%%%%%%%%%%%%%%%%%%%

%%%%%%%%%%%%%%%%%%%%%%%%%%%%%%%%%%%%%%%%%%%%%%%%%%%%%%%%%%%%%%%
\section{Results}\label{sec:results}
%%%%%%%%%%%%%%%%%%%%%%%%%%%%%%%%%%%%%%%%%%%%%%%%%%%%%%%%%%%%%%%

In \figref{fig:line_plot}, we show the test statistic as a function of the C$\nu$B overabundance parameter $\eta$. 
Since the SM cross-sections are proportional to the centre-of-mass energy $s$, and because we assume the relic neutrinos to be at rest, the values of $\eta$ that are probed will depend on the neutrino masses. 
As the absolute scale of the neutrino masses is not known, this analysis was repeated with different assumptions on the mass of the lightest neutrino, assuming normal ordering (NO) and using the mixing best-fit parameters from the NuFit global analysis~\cite{NuFIT:2022}.

We consider three scenarios for the value of the lightest neutrino mass.
First, we employ the constraints on the sum of the neutrino masses coming from cosmological measurements, i.e. $\sum_i m_i < 0.13$ eV \cite{DES:2023}, and 
the best-fit values of the mass splittings from neutrino oscillation experiments to obtain a mass for the lightest neutrino of $m_1 = 0.0342$\,eV. 
Direct searches for neutrino masses, such as the one carried away by KATRIN experiment, set bound on the effective electron neutrino mass of $\abs{U_{ei}}^2 m_i^2 < 0.8\, \mathrm{eV}^2$ \cite{KATRIN:2022}, which is then translated to the value of the lightest neutrino mass of $m_1 = 0.8$~eV. 
Finally, we also consider a case where the lightest neutrino mass is small, but $\nu_1$ is still non-relativistic today, i.e. $m_1 = 0.001$~eV. 

The strongest constraints on $\eta$ come from the larger values of $m_1$, resulting from the increased centre-of-mass energy of the scattering processes, which leads to stronger interactions between the neutrinos. The limiting factor on the strength of the constraint, i.e. the plateau observed at higher values of $\eta$, is due to the limited strength of our analysis of NGC 1068 as a point source. At these values, the signal pdf of the scattered neutrino model goes to zero for all events as no signal events are predicted to be detected; this, in turn, means that the model likelihood tends towards a pure background. To push the exclusion bounds to higher confidence levels (C.L.), we would require a larger likelihood value for the PL point-source analysis of NGC 1068.

\begin{figure}[t!]
    \centering
    \includegraphics[width=\linewidth]{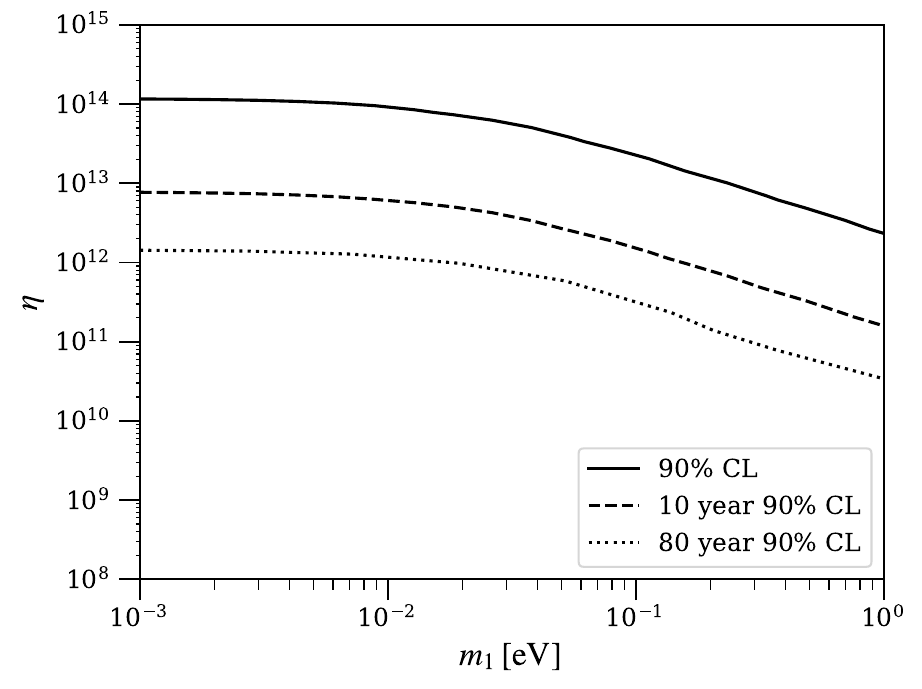}
    \caption{
    The combined 90$\%$ confidence limits from all three sources on a global relic overabundance as a function of the mass of the lightest neutrino. The solid line is the constraint using current data, and the dashed(dotted) line shows the expectation of the limit with an extra 10(80) years of data taking.
    We assume normal ordering and take the mass splittings from NuFIT 2022 \cite{NuFIT:2022}. 
    }
    \label{fig:2d_plot}
\end{figure}

Extending this analysis to a range of values of $m_1$ gives the results shown in \figref{fig:2d_plot}, where the 90\% confidence limits are plotted for current data, and the expected sensitivity for an extra 10 and 80 years of data taking using the methods explained previously. We choose 80 years as a proxy for 10 years of data taking at the future IceCube Gen 2 experiment. Our constraints improve as $m_1$ increases due to the increase in the centre-of-mass energy. On the other hand, for small $m_1$, the mass squared differences dominate in setting the mass scale, which results in an asymptotic limit. 
%%%%%%%%%%%%%%%%%%%%%%%%%%%%%%%%%%%%%%%%%%%%%%%%
%%%%%%%%%%%%%%%%%%%%%%%%%%%%%%%%%%%%%%%%%%%%%%%%

\begin{figure}[t!]
    \centering
    \includegraphics[width=\linewidth]{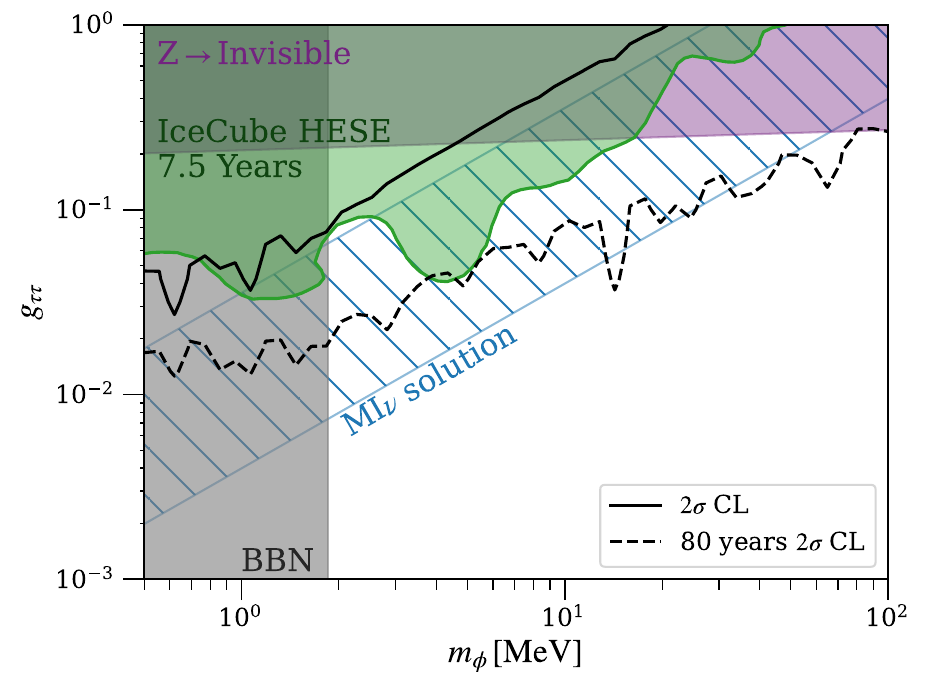}
    \caption{{\bf Scalar mediated neutrino self-interaction.} The $2 \sigma$ exclusion region determined by IceCube is shown based on astrophysical sources NGC 1068, TXS 0506+056 and PKS 1424+240, and their combined result (black solid line). To produce these, we assumed normal ordering of the neutrino masses and took $\sum_i m_i = 0.1 \mathrm{eV}$ with mass splittings from \cite{NuFIT:2022}. Additionally, we include the projected exclusion region for an additional 80 years of operation (black dash line), equivalent to 10 years of data collection with IceCube-Gen2. We compare our results to those from \cite{Esteban:2021}, which includes exclusion regions from BBN (grey), Z to invisibles decay (purple), and IceCube HESE (green).}
    \label{fig:2d_plot_SI}
\end{figure}

The results of our analysis of NSI are presented in \figref{fig:2d_plot_SI}. We compare our bounds to those from BBN, Z to invisible decays, and High Energy Starting Events at IceCube \cite{Esteban:2021}. We find that the combined bound from all three sources considered in this work is generally outperformed in the relevant parameter space by the diffuse source analysis, which we attribute to the smaller number of source events for point-sources. However, the bounds are comparable, and point sources provide better constraints for mediator masses, $m_{\phi}$,  below 1 MeV, though BBN already constrains this. These bounds are both complementary since calculating the flux of diffuse sources in the NSI scenario requires assumptions about the $z$-dependence of neutrino production in the early Universe \cite{Esteban:2021}, which is not the case for point sources where the redshift is known to within some uncertainty. We have also performed the analysis for 80 years of extra data as a proxy for 10 years of IceCube Gen-2 (dashed line in \figref{fig:2d_plot_SI}). The increase in observed events expected at IceCube Gen-2 will be able to push the exclusion region to cover much of the MI$\nu$ solution, which will have important consequences for both neutrino physics and cosmology. New analysis techniques and reconstruction algorithms may further improve this.

It is also possible that other BSM scenarios could be constrained using a similar analysis. We performed a log-likelihood analysis on neutrino decays arising from a coupling to a massless Majoron. However, we found that current public data cannot constrain this scenario to a statistically significant degree as $\abs{TS} < 0.2$ for all coupling values. This may change with improved reconstruction and more data \cite{Valera:2023bud}.
%%%%%%%%%%%%%%%%%%%%%%%%%%%%%%%%%%%%%%%%%%%%%%%%
%%%%%%%%%%%%%%%%%%%%%%%%%%%%%%%%%%%%%%%%%%%%%%%%

%%%%%%%%%%%%%%%%%%%%%%%%%%%%%%%%%%%%%%%%%%%%%%%%%%%%%%%%%%%%%%%
\section{Final Thoughts}\label{sec:conc}
%%%%%%%%%%%%%%%%%%%%%%%%%%%%%%%%%%%%%%%%%%%%%%%%%%%%%%%%%%%%%%%
Point-sources of neutrinos at IceCube offer a unique perspective into neutrino physics, both within and beyond the Standard Model. As more data is collected, and improvements are made to analysis and reconstruction techniques, these sources will be able to push our understanding of then neutrino to even further limits. They are particularly complimentary to diffuse neutrinos sources, as the distance of propagation can be found using astronomical observations of their source galaxy. This allows for a reduction in the assumptions made in any analysis, which will provide more robust results. Furthermore, the addition of new data may make possible the identification of additional point-source candidates beyond the three discussed in this analysis, which will potentially allow for further improvement on the results presented here. This will almost certainly be the case with the future IceCube Gen2 experiment, which will increase the rate of data taking by almost an order of magnitude.
%%%%%%%%%%%%%%%%%%%%%%%%%%%%%%%%%%%%%%%%%
\acknowledgements
%%%%%%%%%%%%%%%%%%%%%%%%%%%%%%%%%%%%%%%%%
We want to thank Jack Shergold for useful discussions and Carlos A. Arg\"{u}elles and   Martin Bauer for reading the draft version of this paper and for their helpful comments.  The UK Science and Technology Facilities Council (STFC) has funded this work under grant ST/T001011/1. 
This project has received funding/support from the European Union's Horizon 2020 research and innovation programme under the Marie Sk\l{}odowska-Curie grant agreement No 860881-HIDDeN.
This work has made use of the Hamilton HPC Service of Durham University. This work used the DiRAC@Durham facility managed by the Institute for Computational Cosmology on behalf of the STFC DiRAC HPC Facility
(www.dirac.ac.uk), which is part of the National eInfrastructure and funded by BEIS capital funding via
STFC capital grants ST/P002293/1, ST/R002371/1 and
ST/S002502/1, Durham University and STFC operations
grant ST/R000832/1.\\

\paragraph*{Note added} After the first version of this manuscript was posted on the arXiv, an independent study by Herrera \textit{et al.} appeared \cite{Herrera:2024upj}. Exploiting the non-observation of the diffuse flux of relic neutrinos expected from interactions of ultra-high-energy cosmic rays with the cosmic-neutrino background, they obtain an upper limit of $\delta \lesssim 10^{4}$ on the \emph{average} $\mathrm{C}\nu\mathrm{B}$ overdensity on cosmological (Gpc) scales. This global bound is complementary to the local-scale constraint presented in the present work.

%%%%%%%%%%%%%%%%%%%%%%%%%%%%%%%%%%%%%%%%%%%%%%%%%%%%%%%%%%%%%%%%%%%%%%%%%%%%%%%%%%
\bibliography{references}
%%%%%%%%%%%%%%%%%%%%%%%%%%%%%%%%%%%%%%%%%%%%%%%%%%%%%%%%%%%%%%%%%%%%%%%%%%%%%%%%%%
\onecolumngrid
\appendix

%%%%%%%%%%%%%%%%%%%%%%%%%%%%%%%%%%%%%%%%%%%%%%%%%%%%%%%%%%%%%%%%%%%%%%%%%%%%%%%%%%
\section{Numerically Solving The Neutrino Transport Equation}\label{sec:SolvingTransportEq}
%%%%%%%%%%%%%%%%%%%%%%%%%%%%%%%%%%%%%%%%%%%%%%%%%%%%%%%%%%%%%%%%%%%%%%%%%%%%%%%%%%
\noindent To produce the flux of muon neutrinos at Earth, it is necessary to solve \equaref{eq:CnuB_transport} for the three mass states $i=1,2,3$. We use a similar method to \cite{Esteban:2021} to discretise the fluxes into bins of energy, and then solve over distance/redshift using an implicit finite difference method. Doing so provides a numerically stable solution to the transport equation.

\subsection*{Upper And Lower Bounds On Neutrino Energy}
\noindent Before describing the details of the numerical method, we will explain the bounds of the neutrino energy integral in the second term of Eq.~\ref{eq:CnuB_transport}. The lower bound, somewhat trivially, is the lowest energy of neutrinos that can be detected by IceCube, which we take to be $E_\mathrm{min} = 100\,$GeV. We can do this because, as seen from the integral in \eqref{eq:CnuB_transport}, the flux at a specific energy depends only on the flux at higher energies. This results in the neutrinos ``flowing down'' in energy. As such, there is no dependence on undetectable neutrinos. \\

\noindent Physically, the upper bound on the neutrino energy is infinity; however, when solving this integral numerically, we need a finite upper bound which will approximate the integral well. To find a value for the finite upper bound, it is useful to look at the integral of the initial power-law flux from this lower bound up to some energy cutoff $E$:
\begin{equation}
    I(E) = \int_{100\,\mathrm{GeV}}^E \Phi(0,E') dE' = \frac{\Phi_0}{1-\gamma} \frac{1}{E_0^{-\gamma}} \left( E^{-\gamma+1} - \left(100\,\mathrm{GeV}\right)^{-\gamma+1}\right)\,,
\end{equation}
where the initial flux is a power law:
\begin{equation}
    \Phi(0,E) = \Phi_0 \left(\frac{E}{E_0}\right)^{-\gamma}\,,
\end{equation}
and we have assumed that $\gamma > 1$ to ensure that the integral converges when $E \rightarrow \infty$. The fraction of the total flux above the cutoff is then given by:
\begin{equation}\label{eq:Flux_fraction}
    f(E) = 1-\frac{I(E)}{I(\infty)} = \left(\frac{E}{10^2\,\mathrm{GeV}}\right)^{-\gamma+1}\,.
\end{equation}
Since the total flux above our cutoff energy is always greater than or equal to the scattered flux above the cutoff energy, this fraction is then an upper bound on the error in the approximation of the integral in \eqref{eq:CnuB_transport}. If we want the fraction of the total flux above our cutoff to be smaller than some $\epsilon$, we can find the corresponding cutoff $E_\mathrm{max}$ from rearranging \eqref{eq:Flux_fraction}. This gives the relation:
\begin{equation}
    \log_{10}\left(\frac{E_\mathrm{max}}{1 \,\mathrm{GeV}}\right) = 2 + \frac{\log_{10}(\epsilon)}{\gamma-1}\,.
\end{equation}
For example, if we want to limit the flux ignored to $\epsilon = 10^{-6}$ we require $E_\mathrm{max} = 10^8$ GeV assuming $\gamma = 2$. In practice, we take a fixed value of $E_\mathrm{max} = 10^{10}$ GeV, which satisfies $\epsilon \leq 10^{-6}$ for $\gamma \geq 1.5$. 

\subsection*{Solving The Transport Equation Without Redshift}
\noindent There are two cases in which we need to solve the transport equation; for closer sources where redshift effects can be ignored, and for further sources where the redshift effects need to be accounted for. The former of these is a simpler case, and so we will first describe the discretisation and solving of the equation before adapting the algorithm for the latter case.\\

\noindent To ignore the effect of redshift, we can set the energy loss term in \equaref{eq:CnuB_transport} to zero. We can also use the fact that the neutrinos are relativistic and change variables from time $t$ to distance $r$. We then get:
\begin{equation}
   \frac{\partial \Phi_i(r, E)}{\partial r} = -\Phi_i(r, E)\sum_j n_j \sigma_{ij}(E) + \sum_{j,k,l}n_k\int_E^\infty dE^\prime \Phi_j(r,E^\prime)\frac{d \sigma_{jk\rightarrow il}(E^\prime,E)}{dE} 
\end{equation}
This can be written more efficiently by assuming that all neutrino mass states have the same relic density, i.e. $n_i = n_\nu \, \forall \, i \in \left\{1,2,3 \right\}$, and repackaging the sums over integrated and differential cross-sections.
\begin{equation}\label{eq:transp_final}
    \frac{1}{n}\frac{\partial \Phi_i (r, E)}{\partial r} = - \Phi_i (r,E) K_i(E) + \sum_{j}\int_E^\infty dE' \ \Phi_j(r,E') J_{ji}(E,E')
\end{equation}
where the function $K_i$ contains the cross-sections for incoming neutrino with mass $m_i$, and the kernel function $J_{ji}$ contains the differential cross-sections for the incoming neutrino with mass $j$ and outgoing neutrino with mass state $i$. The sum over $k$ and $l$ has been moved inside of $J_{ji}$. 
To discretise the flux over energy, we project the differential equation onto a set of basis functions given by $\{ \Theta(E-E_{n-1/2})\Theta(E_{n+1/2}-E) | n < N, \, n,N \in \mathbb{N}\}$, where $\Theta(x)$ is the Heaviside step function and N is the number of basis functions. This improves the numerical stability of the solution when there are discontinuities, such as those in the cross-sections which arise from $e^+e^-$ production, compared to a finite-difference based discretisation. In our implementation, we space the bins logarithmically and take the total number of bins $N$ as 300. We then integrate the differential equation over energy to obtain $N$ coupled equations:
\begin{equation}\label{eq:energy_discretised}
    \frac{\Delta E_n}{n_\nu}\parc{\Phi_i^n(r)}{r} = -\Phi_i^n(r) K_i^n + \sum_j\sum_m\Phi_j^m J_{ji}^{m n}\,,
\end{equation}
where we have defined:
\begin{align*}
    \Phi_i^n(r) &= \frac{1}{\Delta E_n}\int_{E_{n-1/2}}^{E_{n+1/2}} dE \ \Phi(r,E) \\
    K_i^n &= \int_{E_{n-1/2}}^{E_{n+1/2}} dE\ K_i(r, E) \\
    J_{ji}^{mn} &= \int_{E_{n-1/2}}^{E_{n+1/2}} dE  \int_{E_{m-1/2}}^{E_{m+1/2}} dE^\prime \ J_{ji}(E, E^\prime),&&\text{ if } m > n \\
    J_{ji}^{mn} &= \int_{E_{n-1/2}}^{E_{n+1/2}} dE \int_{E}^{E_{m+1/2}} dE^\prime \ J_{ji}(E, E^\prime),&&\text{ if } m=n \\
    J_{ji}^{mn} &= 0 ,&&\text{ if } m<n  \\
    \Delta E_m &= E_{m+1/2}-E_{m-1/2} \,.
\end{align*}
There are three possible values of $J_{ji}^{mn}$, each corresponding to different cases. First, we have that the $m$th energy bin is higher than the $n$th (or equivalently $m > n$). In this case we can integrate over both energy limits independently. In the second case however, we have $m = n$ which implies that the energy bins are the same. Since the initial energy $E'$ must be greater than the final energy $E$, the lower energy limit in the second integral is $E$ rather than $E_{m-1/2}$. Finally, we have the case of $m<n$ which is not possible as the energy must decrease, resulting in a value of zero. We calculate the integrals of $K_i^n$ and $J_{ji}^{mn}$ analytically, which reduces the time needed to solve the equation. \\

\noindent We now discretise the distance $r$, following an implicit finite-difference scheme. This amounts to the substitutions:
\begin{align}
    \frac{\partial \Phi_i^n (r)}{\partial r} &\rightarrow \frac{\Phi_i^n(r_{a+1}) - \Phi_i^n(r_a)}{\Delta r} \\
    \Phi_i^n(r) &\rightarrow \Phi_i^n(r_{a+1})
\end{align}
Since $K$ and $J$ are not functions of $r$, this discretisation does not affect them. Performing these substitutions in Eq.~\eqref{eq:energy_discretised} gives:
\begin{equation}\label{eq:fully_discretised}
    \frac{\Delta E_n}{n_\nu} \frac{\Phi_i^n(r_{a+1}) - \Phi_i^n(r_a)}{\Delta r} = - \Phi_i^n(r_{a+1}) K_i^n + \sum_j\sum_m \Phi_i^m(r_{a+1}) J_{ji}^{mn}\,.
\end{equation}
We now have a fully discretised form of \equaref{eq:CnuB_transport}, which we need to solve starting from our initial power-law flux.\\

\noindent To solve these equations, we follow a similar method to that used in \cite{Esteban:2021}. We rewrite \equaref{eq:fully_discretised} as a matrix equation over the $i$ and $j$ indices:
\begin{equation}\label{eq:matrix_eqn}
    M_{ji} x_j = \frac{x^\prime_i}{\Delta r} - \frac{n_\nu y_i}{\Delta E_n} \,,
\end{equation}
where
\begin{align}
    x_i &= \Phi_i^n(r_{a+1}) \\
    x^\prime_i &= \Phi_i^n(r_a) \\
    y_i &= \sum_{m>n}\sum_j \Phi^m_i(r_{a+1}) J^{mn}_{ji} \\
    M_{ji} &= \left(\frac{1}{\Delta r} + \frac{n_\nu}{\Delta E_n} K_i^n\right)\delta_{ij} - n_\nu J^{nn}_{ji} \\
\end{align}
Since each equation for $n$ depends only on the solutions of equations $m>n$, we can solve the whole system by starting at $n=N$ and propagate the solutions down to $n=0$. This way, the only unknowns are the values $x_j$, which can be solved for using standard linear algebra techniques.\\

\noindent  One final step is to introduce a lower bound value for the flux in a bin, below which it is set to zero. Doing so reduces noise and improves the stability of the solution dramatically. The cutoff value was found by trial and error not to lose any useful information about the flux. We found that a value of $\Phi_\mathrm{min} = 10^{-30}\,\Phi_0$ was sufficient, as it provided a stable solution without truncating the flux at too low an energy. Since the output of the numerical solver was used to calculate a pdf, the normalisation $\Phi_0$ does not matter, so we set it to be 1.

\subsection*{Solving The Transport Equation With Redshift}
\noindent  Solving the transport equation whilst taking into account redshift can be done in a similar method to the one previously described, with some modifications.\\

\noindent Firstly, we wish to solve the flux as a function of the redshift $z$, using the substitution $\partial_t = H(z) (1+z) \partial_z$. Note that the definition of energy in the interaction is in the local frame of reference, i.e. $E_\mathrm{local} = (1+z)E_\mathrm{observed}$. In \equaref{eq:CnuB_transport}, this fact is taken into account by the first term, however in our discretised equation we can instead choose the energy bins such that we can make the substitution
\begin{equation}
    \Phi^n_i(z_{a}) \rightarrow \left(\frac{1+z_a}{1+z_{a+1}}\right)^{-3}\Phi^{n+1}_i(z_{a})\,.
\end{equation}
This means that the effect of redshift on the energy will shift the bins down by one for each step in redshift, and to also dilute the flux due to the expansion of the Universe. In order for the former to be the case, we have the following relation:
\begin{equation}
    -\log_{10}\left(\frac{1+z_{a+1}}{1+z_a}\right) = \Delta\log_{10}E\,,
\end{equation}
where the minus sign is needed due to the value of $z$ starting at some $z_\mathrm{max} > 0$ and decreasing monotonically to $z = 0$ as the neutrinos propagate, i.e. $z_{a+1} < z_a$. The values of $z$ for which we solve can then be defined iteratively:
\begin{equation}
    z_{a+1} = \frac{1+z_a}{\Delta\log_{10}E} - 1 \,.
\end{equation}
To find the number of steps in $z$ needed, we use the fact that
\begin{equation}
    1+z_\mathrm{max} = \prod_{a=0}^{N_z - 1}\frac{1+z_{a}}{1+z_{a+1}} \,,
\end{equation}
which gives
\begin{equation}
    N_z = \frac{\log_{10}(1+z_\mathrm{max})}{\Delta\log_{10}E}\,.
\end{equation}
Following this procedure will properly account for the effect of redshift on the flux as it propagates through the expanding Universe to Earth \cite{Lee:1996fp}.

\noindent There is one additional effect which differs when taking redshift into account, which is the fact that the relic density of neutrinos will also decrease as the redshift also decreases. This is easily included by promoting $n_\nu$ to a function of $z$, writing it in terms of the current neutrino density:
\begin{equation}
    n_\nu(z) = \left(1+z_{a+1}\right)^3 56 \,\mathrm{cm}^{-3}\,.
\end{equation}
Combining all these effects, and combining with the energy discretisation outlined in the previous section, leaves us with the following set of equations:
\begin{equation}
    \bm{M}_{ji} \Phi_j^n(z_{a+1}) = \frac{H(z) (1 + z_{a+1})}{z_a - z_{a+1}} \left(\frac{1+z_a}{1+z_{a+1}}\right)^{-3}\Phi^{n+1}_i(z_a) - \frac{n_\nu(z)}{\Delta E_n} y_i \,,
\end{equation}
where
\begin{align}
    y_i &= \sum_{m>n}\sum_j \Phi^m_j(z_{a+1})J_{ji}^{mn}\,, \\
    M_{ji} &= \left( \frac{H(z)(1+z_{a+1})}{z_a - z_{a+1}} + \frac{n_\nu(z)}{\Delta E_n}K_i^n \right) \delta_ij - n_\nu(z)J^{nn}_{ji}\,.
\end{align}
This system of equations can be solved using the same method as previously described.

\section{Standard Model Interactions}\label{sec:CrossSections}

\begin{figure}[h!]
\captionsetup[subfigure]{labelformat=empty}
\centering
\subfloat[\label{fig:t-channel_Z}]{}
\subfloat[\label{fig:u-channel_Z}]{}
\subfloat[\label{fig:s-channel}]{
\centering
    \includegraphics{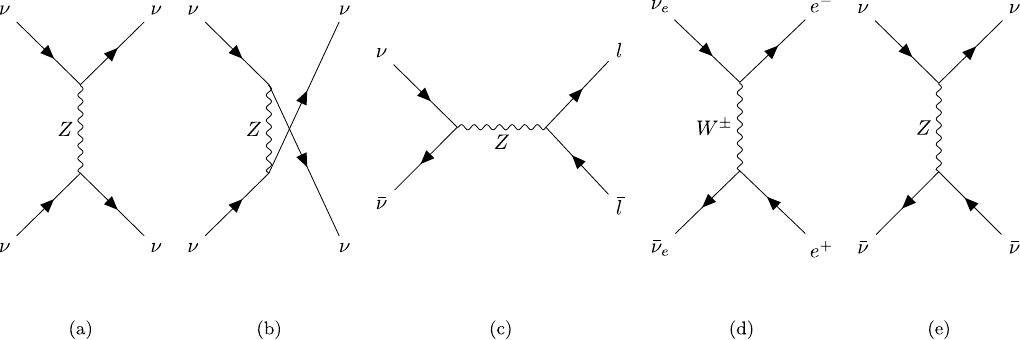}
}
\subfloat[\label{fig:t-channel_W}]{}
\subfloat[\label{fig:t-channel_Z_anti}]{}
\caption{Feynman diagrams for the relevant interactions between neutrinos and relic neutrinos. (a) and (b) are the $t$- and $u$-channel diagrams, respectively, for elastic scattering of incident neutrinos off of relic neutrinos. (c) is for lepton pair production where $l = \nu$ or $l=e^-$. (d) is the $t$-channel $e^+e^-$ pair production diagram. (e) is the $t$-channel elastic scattering of neutrinos on relic antineutrinos.}
\end{figure}
\subsection*{Total cross-sections}
\noindent  We assume that all particles in the incoming flux are neutrinos rather than antineutrinos. This is valid as the flux observed at IceCube is a combination of $\nu_\mu$ and $\Bar{\nu}_\mu$ fluxes, and the total cross-section is invariant under swapping $\nu \leftrightarrow \bar{\nu}$. \\

\noindent We start with the $\nu\nu \rightarrow \nu\nu$ scattering. Since the $ Z$-boson mediates this process, it is convenient to work in the neutrino mass basis. The total cross-section is:
\begin{equation}
    \sigma_{ij} = \frac{G_F^2 s_j (3\delta_{ij} + 1)}{2\pi} \,,
\end{equation}
where $G_F$  is Fermi's constant and  the $3\delta_{ij}$ occurs because of the interference between $t$ and $u$-channel diagrams, as shown in \figref{fig:t-channel_Z} and \figref{fig:u-channel_Z} respectively, which occurs when $i = j$. In the case of $\nu \Bar{\nu}$ scattering, we separate the cross-section calculation into two categories - $\nu\Bar{\nu}$ and $e^+e^-$ production. The first of these follows similarly to the previous case, in particular when $i \neq j$:
\begin{equation}
    \sigma_{ij} = \frac{G_F^2 s_i}{6\pi} \,,
\end{equation}
while for $i=j$, we have to account for the annihilation and production of new neutrino pairs. Combined, this gives:
\begin{equation}
    \sigma_{ii} = \frac{G_F^2 s_j}{\pi}\left(\frac{2}{3} + 2\times\frac{1}{6} \right) = \frac{G_F^2 s_j}{\pi}\,,
\end{equation}
where the first term arises from production on a $\nu_i \Bar{\nu}_i$ pair, which receives an enhancement from the additional $t$-channel diagram as shown in \figref{fig:t-channel_Z_anti}. The second term is from the production of a $k$ state mass pair, where $i \neq k$, of which there are two possibilities. \\

\noindent For the electron pair production process, the cross-section must be calculated on a weak basis. We write the mass-basis cross-section in terms of the weak basis cross-section using the PMNS matrix:
\begin{equation}
    \sigma_{ij} = \sum_{\alpha,\beta} \abs{U_{i\alpha}}^2 \abs{U_{j\beta}}^2 \sigma_{\alpha\beta}\,.
\end{equation}
\begin{equation}
  \sigma_{\alpha\alpha} = \frac{G^2_F}{3 \pi}{\sqrt{1-\frac{4 m^2_{e}}{s}}\left(2 m^2_{e}\left(g^2_{V} -2g^2_{A}\right)+s\left(g_{A}^2+g_{V}^2\right)\right)}
  \end{equation}
  where $g_{A} = \delta_{\alpha e} + g^{\alpha}_{A}$ and $g_{V} = \delta_{\alpha e} + g^{\alpha}_{V}$ where $g^{\alpha}_{A} = -\frac{1}{2}$ and $g^{\alpha}_{V} = -\frac{1}{2} +2\sin^2\theta_{W}$.
  We note that this includes all relevant flavour combinations. 

\subsection*{Differential cross-sections}
\noindent  For the process $\nu_j\nu_k\rightarrow\nu_i\nu_l$ we find that:
\begin{equation}
    \der{\sigma_{jk\rightarrow il}}{t} = \frac{G_F^2}{4\pi}(\delta_{ij}\delta_{kl}+\delta_{ik}\delta_{jl})^2\,,
\end{equation}
where the Mandelstam variables are $t = -2m_k(E^\prime-E)$ and $u = -2m_kE$. The process $\nu_j\Bar\nu_k\rightarrow\nu_i\Bar\nu_l$ follows similarly, with the Mandelstam variables remaining the same:
\begin{equation}
    \der{\sigma_{jk\rightarrow il}}{t} = \frac{G_F^2}{4\pi}\frac{u^2}{s_k^2}(\delta_{jk}\delta_{il}+\delta_{ij}\delta_{kl})^2\,.
\end{equation}
Finally, we also need to account for up-scattered relic antineutrinos, i.e. the process $\nu_j\Bar\nu_k\rightarrow\Bar\nu_i\nu_l$. This differs from the previous two cases as the Mandelstam variables $u$ and $t$ are swapped. The differential cross-section for this process is then:
\begin{equation}
    \der{\sigma_{jk\rightarrow il}}{t} = \frac{G_F^2}{4\pi}\frac{t^2}{s_k^2}(\delta_{jk}\delta_{il}+\delta_{ij}\delta_{kl})^2\,.
\end{equation}
Combining these different processes gives the final differential cross-section: 
\begin{equation}
    \frac{d\sigma_{jk\rightarrow il}(E^\prime,E)}{dE} = \frac{G_F^2 m_k}{2\pi}\left((\delta_{ij}\delta_{kl}+\delta_{ik}\delta_{jl})^2 + (\delta_{jk}\delta_{il}+\delta_{ij}\delta_{kl})^2\left(\frac{E}{E^\prime}\right)^2 \right)\,,
\end{equation}

\end{document}